\documentstyle[aps,prl,floats]{revtex}
\begin{document} 
\draft 
\twocolumn[\hsize\textwidth\columnwidth\hsize\csname @twocolumnfalse\endcsname

%\preprint{Preprint } 
\title{Josephson-plasma and vortex modes  in layered superconductors}  
\author{E.B. Sonin}

\address{Racah Institute of Physics, Hebrew University of
Jerusalem, Jerusalem 91904, Israel } 

\date{\today} \maketitle

\begin{abstract}
  The Josephson-plasma and vortex modes in layered  superconductors have been
studied theoretically for low magnetic fields parallel and perpendicular to
layers. The two modes belong to the same lowest-frequency branch of
collective-mode spectrum localized near vortices. This is the
Josephson-plasma resonance if pancakes are strongly pinned and cannot
move. Otherwise the lowest-frequency mode is a vortex mode governed by pancake
pinning and the vortex mass. The latter is strongly enhanced by wandering of a
vortex line around its average direction. The recently observed jump of the
magnetoabsorption-resonance frequency at the vortex phase transition  line is
interpreted as a transition from the Josephson-plasma to the vortex mode.
  \end{abstract}

\pacs{PACS numbers:  74.60.Ge, 72.30.+q, 74.72.Hs, 74.25.Nf}  
%\eject
] 
%\narrowtext
%\twocolumn

Magnetoabsorption microwave resonances observed in Bi compounds
(see \cite{M0,Tam} and references therein) have become one of the most
important subjects for studying the interlayer Josephson coupling. The
resonances were observed for magnetic fields normal and nearly parallel to
superconducting layers, and in the absence of the field. A remarkable
feature of the resonances was that at high magnetic fields $H$  the resonance
frequency decreases roughly as $1/\sqrt{H}$ (the anticyclotronic
behavior). Recently the magnetoabsorption resonances were observed in low perpendicular
magnetic fields $H$ \cite{M0,Tam,MatUnp}. In low fields the
resonance frequency very weakly depends on $H$, but at a field about a few
hundred G, the magnetoabsorption resonance frequency jumps to lower values with
the anticyclotronic dependence, as observed in earlier experiments in higher
fields.

The first attempt to explain these resonances for
high perpendicular magnetic fields was done by Kopnin {\em et al.} \cite{KS} who
related the resonances to a vortex mode governed by surface pinning and the
Magnus force. They were able to explain the anticyclotronic behavior, but the
essential role of surface pinning and the Magnus force was not been confirmed by
further experiments. 

Later the majority of researchers came to conclusion that the magnetoabsorption
resonances are related to the Josephson Plasma Resonances (JPR) (see \cite{BKVM}
and references therein) with the frequency 
\begin{equation}
\omega_0^2 =\omega_p^2\langle \cos \varphi_0(\vec r) \rangle
      \label{JPR}~,       \end{equation}
where $\omega_p$ is the Josephson-plasma frequency at
zero magnetic field, $\vec r$ is the inplane coordinate, and  $\varphi_0(\vec r)$
is the stationary gauge-invariant phase difference between layers, which is
nonzero due to misalignment of pancakes in neighboring layers by thermal
fluctuations and disorder. 

However, this interpretation suffers from a  number of inconsistencies
(see discussion in Refs. \cite{PRL,GMB,Repl,Com}), and the original
idea to relate the magnetoabsorption resonances with  a vortex mode
\cite{KS}  should not be ruled out, though properties of this vortex mode must
be different from those assumed in Ref. \cite{KS}. In Ref. \cite{Com} the  
magnetoabsorption resonances  in high parallel magnetic fields were interpreted
as a vortex mode with the frequency determined by the balance of the pinning
force on pancakes and the inertia force determined by the vortex mass. The
Magnus force is not effective in the parallel geometry.

In the present work I investigate how a growing low magnetic field, either
parallel or perpendicular to layers, changes the collective-mode spectrum in
layered superconductors. The goal is to find correspondence of the
JPR and the vortex mode to branches of this spectrum. In order to avoid a 
semantic ambiguity, I shall use the name JPR only for  the mode with the
frequency given by Eq. (\ref{JPR}), as done  in all recent experimental papers,
though in old papers this name had not the same meaning  (see  below). The
vortex mode is defined as a Goldstone mode which has a finite frequency only due
to pinning. I shall show that the JPR mode and the vortex mode belong to the
same branch of the spectrum which is the lowest-frequency oscillation mode
localized near vortices. This mode becomes the JPR  only if pancakes cannot move
because of very strong  pinning. If the pinning strength is finite,
the lowest-frequency mode is a vortex mode  governed by pinning and the vortex
mass. The vortex mass is related with the electric energy in the interlayer
spacing, and the misalignment of pancakes in the  perpendicular field essentially
increases the vortex mass. On the basis of the presented analysis I suggest that
in the recent magnetoabsorption experiments in low perpendicular magnetic fields
\cite{Tam,MatUnp} they observed a transition from the JPR to the vortex mode
accompanied by a frequency jump.

The analysis will be done for a single long Josephson junction,
but results of the analysis are relevant also  for a layered
superconductor after slight modifications (see below). The state of the junction
is determined from the sine-Gordon equation for the phase:
 \begin{equation}
{1\over \lambda_J^2} \sin \varphi -
 \nabla^2 \varphi =- {1\over c_s^2} \ddot \varphi~, 
   \label{ded} \end{equation}
where $c_s =c\sqrt{s/ 2 \varepsilon\lambda}$ is  the Swihart velocity, $s$ is the
barrier thickness, or the interlayer spacing,  $\lambda$ is the London
penetration depth for bulk superconductors forming the junction ($\lambda \gg
s$), $\varepsilon$ is the dielectric permeability, $\lambda_J^2 ={\Phi_0^2
/32\pi^3 \lambda E_J}$ is the Josephson length, $\Phi_0$ is the magnetic-flux
quantum, and $E_J$ is the Josephson coupling energy. 

The collective modes correspond to small oscillations around the stationary
solution $\varphi_0(x)$: $\varphi(x,t) = \varphi_0(x) + \varphi '(x,t)$. The
small phase $\varphi ' \propto e^{-i\omega t}$ is determined by
the equation (we shall omit the prime later on)  
\begin{equation}
- \nabla^2 \varphi - {1- \cos \varphi_0 \over \lambda_J^2} \varphi
 ={\omega^2 -\omega_p^2 \over c_s^2}\varphi~. 
  \label{osc}  \end{equation}
In the absence of the magnetic field the ground state corresponds to
$\varphi_0=0$, and  the solutions of Eq. (\ref{osc}) are plane waves
$\exp(i\vec k\vec r-i\omega t)$. The oscillation spectrum has a plasma 
edge, i.e.,
$\omega =\sqrt{\omega_p^2 +c_s k^2} > \omega_p$,
where $\omega_p^2 ={e^2 E_J s /\varepsilon \hbar^2} ={c_s^2 / \lambda_J^2}$
 is the  Josephson plasma frequency.

If the magnetic field is strictly
parallel to the junction plane, the
phase $\varphi(x,t)$ depends only on one spatial coordinate $x$. 
In the stationary case ($\dot \varphi =0$) the sine-Gordon equation has 
exact periodic solutions $\varphi_0(x)$ which correspond to chains of Josephson
vortices \cite{Fet}  and the period (intervortex distance) $a$ determines the
magnetic field: $H=\Phi_0/2\lambda a$. For such  $\varphi_0(x)$,  Eq.
(\ref{osc}) can be also solved exactly for any intervortex distance \cite{Fet}. 
However, we introduce  small periodic $\varphi_0(x)$  which is not a
solution of the sine-Gordon equation and vanishes at distances more than 
$\lambda_J/2$ from vortex centers. This yields a small periodic attractive
potential $U=[1- \cos \varphi_0(x)]/\lambda_J^2$ in Eq. (\ref{osc}). We 
assume that  $a \gg \lambda_J$ (low magnetic fields), and look for
a periodic solution  symmetric in the interval $-a/2 <x<a/2$. The coordinate $x=0$ 
corresponds to the center
of the potential well. At $|x| <\lambda_J/2$  the perturbation theory yields
$\varphi(x) \propto 1-\int_0^x dx_1  \int_0^{x_1} dx_2 U(x_2)$. At 
$\lambda_J/2<|x| <a/2$ $\varphi(x) \propto \cos k(a/2 \mp x)$.  The
lowest-frequency mode corresponds to the bound state
 with an imaginary $k=ip$. Continuity of the phase and its derivative at $x
\sim \pm \lambda_J/2 $ requires that \begin{equation}  p \tanh p a =  {1 \over
\lambda_J^2} \int_{-a/2}^{a/2} dx [1- \cos \varphi_0(x)]  ~.
 \label{tan}   \end{equation}
At very low magnetic fields when $pa\gg 1$, 
the bound state corresponds to the oscillation frequency
\begin{equation}
\omega_0^2=\omega_{p}^{2}-c_{s}^{2}p^{2} =\omega_p\left\{1- {a^2 \over
\lambda_J^2} \left[1-\langle \cos \varphi_0(x) \rangle \right]^2 \right\}~.
             \end{equation}
But if  the inequality $pa \ll 1$ holds, expansion in the left-hand side of
Eq. (\ref{tan}) yields
\begin{equation}
p^2 \approx  {1 \over a \lambda_J^2} \int_{-a/2}^{a/2} dx [1- \cos \varphi_0(x)]
={1- \langle \cos \varphi_0(x) \rangle \over \lambda_J^2}~,
    \end{equation}
and the squared mode frequency $\omega_0^2$ is given by
Eq. (\ref{JPR}).

Thus the JPR belongs to the lowest-frequency mode localized near the
potential well.  However, the potential is very weak, the size of the
bound state $1/p$ (the localization length) is larger than the intervortex
distance $a$, and therefore localization is not so pronounced. 

Let us consider also the lowest-frequency oscillation which belongs
to the continuum spectrum  with a real $k$. The phase distribution for a
continuum mode must have at least one node in the vicinity of the vortex, in
order to satisfy the condition of orthogonality to the ground bound state. Thus
roughly $k= \pi/a$, and the lowest frequency $\omega_c$ of the continuum
spectrum (delocalized phase oscillation)   exceeds the zero-field plasma edge
$\omega_p$:  
\begin{equation}
\omega_c^2=\omega_{p}^{2}+c_{s}^{2}k^{2} =\omega_p^2\left(1 +{\pi^2 \lambda_J^2
\over a^2}\right)~.
 \label{cont}   \end{equation}
The  frequency of this mode grows at increasing the magnetic field. Note that
in the old papers \cite{Fet} this mode with $\omega_{c} >\omega_p$, but not the
localized mode with $\omega_0 < \omega_p$, was called the plasma mode. 

Our calculation has not revealed the Goldstone vortex mode, because our
potential does not correspond to a real Josephson vortex with the phase
$\varphi_0$ which satisfies the stationary sine-Gordon equation. For 
real vortices the
perturbation theory is invalid, but the exact theory yields
that the lowest frequency $\omega_0$  vanishes. The phase in the bound
state is proportional to the derivative of the stationary phase distribution
$d\varphi_0(x)/dx$.

So the procedure to tune the
potential  from zero to values of the order $1/ \lambda_J^2$ goes through
states which are not realized physically, since the vortex is a topological
nonlinear excitation, and its topological charge cannot change continuously.
However, this procedure clearly demonstrates genesis of the JPR and the vortex
mode from the original spectrum of the spatially uniform state. The JPR and
the vortex mode belong to the same spectrum branch related to a phase
oscillation localized near the defect (vortex), but the JPR with the frequency
Eq. (\ref{JPR}) is possible only for an artificial weak potential. Thus in
parallel fields {\em the JPR mode doesnot exist at all} \cite{PRL}.

In a layered superconductor a perpendicular field creates vortices which are
chains of pancakes. If pancakes lie on an ideally straight line normal to
layers, there is no phase difference across Josephson junctions. But
because of disorder and thermal fluctuations, pancakes do not form an ideally
straight line, and pancakes in neighboring layers are separated by 
random distances  \cite{LG}. For a single Josephson junction this means that the
vortex line in two banks of the junction meets the junction plane in two
different points with the distance $r_w$ between them, and a phase difference
$\varphi_0(\vec r)$ across the junction appears. If $r_w \gg \lambda_J$, the
phase distribution corresponds to {\em a Josephson string}, which is a segment 
of a Josephson vortex of width $\lambda_J$ stretched between the points
($x=0,y=0$) and ($x=0,y=r_w$). If $r_w \ll \lambda_J$, the statinary
sine-Gordon equation yields that at $r \ll \lambda_J$ $\varphi_0(\vec r)$ is a
dipole field:   
\begin{equation}
\varphi_0(\vec r) = \arctan{y \over x} - \arctan {y-r_w\over x}~.
       \end{equation}
At large distances $r=\sqrt{x^2+y^2} \gg r_w$
(but not larger than $\lambda_J$) the phase is $\varphi_0(\vec r) = r_w
x/r^2$. At distances $r \gg \lambda_J$ the phase $\varphi_0(\vec r)$ is
exponentially small. We shall call such a phase distribution {\em a short
Josephson string}.

We look for the lowest-frequency mode for  a short Josephson string assuming for
the sake of simplicity that the potential $-(1-\cos \varphi_0) /\lambda_J^2$ in 
Eq. (\ref{osc}) is  axisymmetric. The most important perturbation originates from
distances $r_w \ll r \ll \lambda_J$ where the axisymmetric part of $1-
\cos\varphi_0(r) \approx \varphi_0(r)^2/2$ is equal to $r_w^2/4r^2$. An
axisymmetric mode is determined from the equation 
\begin{equation}
- {1\over r}{d\over dr}\left(r{d \varphi\over dr} \right)  - {r_w^2 \over
4\lambda_J^2} {1\over r^2} \varphi
 ={\omega^2 -\omega_p^2 \over c_s^2}\varphi~. 
 \label{e2}   \end{equation}
For a bound state in a weak potential one may neglect the right-hand side of
Eq. (\ref{e2}) for $r<\lambda_J$, and  the perturbation theory yields
\begin{equation}
\varphi(r) = 1 -{r_w^2 \over 8\lambda_J^2}\left(\ln{r\over r_w} \right)^2~. 
   \label{i}    \end{equation}

Outside the potential well, at $r > \lambda_J$, the axisymmetric bound-state
oscillation is $\varphi =A I_0(pr)  +B K_0(pr)$, or, if $pa \ll 1$,
$\varphi \approx  A \left(1+{p^2r^2/
4}\right) -B \ln{ pr}$. The periodical boundary conditions in a real vortex array can be approximately
simulated by the condition $\varphi'(r) =0$ at the
boundary of the Wigner-Zeitz cell $r=a=\sqrt{\Phi_0/\pi H}$. Then
\begin{equation}
\varphi  \approx
A\left( 1+ {p^2a^2\over 2} \ln{1\over pr}\right)~.
    \end{equation}
The continuity conditions at the border of the potential well
($r=\lambda_J$) yield
\begin{equation}
{r_w^2\over 2 \lambda_J^2} \ln{\lambda_J
\over r_w} ={p^2a^2\over 1+ {p^2a^2\over 2} \ln {1\over p\lambda_J}}~.
  \label{bsl}  \end{equation}
In very low magnetic fields ($a \rightarrow \infty$)
\begin{equation}
p \approx
{1\over \lambda_{J} }\exp \left[-{4 \lambda_J^2 \over r_w^2}{1\over
\ln(\lambda_J/r_w)}\right]~. 
    \end{equation}
Then the squared frequency $\omega_{0}^2 =\omega_p^2 -c_s^2 p^2$ is
\begin{equation}
\omega_{0}^2 = \omega_p^2 \left\{1- \exp
\left[-{8 \lambda_J^2 \over r_w^2}{1\over
\ln(\lambda_J/r_w)}\right] \right\}~.
    \label{bnd}   \end{equation}

For a  long Josephson string $r_w \gg \lambda_J$ one maynot use the
perturbation theory, but the lowest-frequency mode can be calculated as a bending
oscillation for an elastic string of  the  length $r_w$ and width $\sim
\lambda_J$ with its ends fixed. The line tension of the string (the string
energy per unit length)  is $\epsilon=8 E_J\lambda_J$.  The mass of the moving
string is determined by the electric energy: 
\begin{equation}
M={2s \over v_{L}^{2}}\int 
{\varepsilon E^2\over 8\pi}(d\vec r)_2
={\varepsilon \Phi_0^2\over 16\pi^3 c^2 s}\int 
\left(\partial \varphi_0
\over \partial x\right)^2(d\vec r)_2~.
   \label{mass} \end{equation}
where $E=(\hbar/2es)\dot \varphi=-(\hbar/2es)(\vec v_L \cdot \vec \nabla) \varphi_0$
 is the electric field parallel to the $c$ axis, and $\vec v_L$ is
the string velocity. Neglecting the logarithmic contribution from vicinities of
the string ends, this expression yields for the mass per unit length 
$\mu_l=M/r_w=\Phi_0^2/2\pi^3 c^2s \lambda_J $ \cite{mass}. Finally, the
frequency is $\omega_0^2={\pi^2 \epsilon/ \mu_l r_w^2} 
=\pi^2 \omega_p^2{\lambda_J^2 / r_w^2}=\pi^2c_s^2/ r_w^2$. At $r_w \sim
\lambda_J$  where the crossover from a short to a long Josephson string takes
place, this frequency is of the same order as given by Eq. (\ref{bnd}). 

We see that low magnetic fields do not affect the lowest-frequency localized mode. But
at increasing $H$, i.e., decreasing the parameter $pa$ in Eq. (\ref{bsl}), the
logarithmic term in the denominator becomes unimportant. Then  the squared
frequency is given by Eq. (\ref{JPR}), with 
\begin{equation}
1-\langle \cos \varphi_0(x) \rangle \approx {\langle\varphi_0(x)^2 \rangle
\over 2}
 = {r_w^2\over 2 a^2} \ln{\lambda_J
\over r_w}~,
\label{phi} \end{equation}
and the bound state corresponds to the JPR mode. 
The restriction on observation of the JPR mode is 
\begin{equation}
p^2a^2  \ln {1\over p\lambda_J}= {r_w^2\over 2 \lambda_J^2}
\ln{\lambda_J\over r_w } \ln{a^2 \over r_w ^2\ln(\lambda_J/r_w)} \ll 1~.
   \label{isl} \end{equation}
Thus the JPR is a weakly localized mode with imaginary $k$,
contrary to Bulaevskii {\em et al.} \cite{BKVM} who related the JPR to the most
homogeneous {\em delocalized} state. Delocalized states belong to the continuum
spectrum, and as well as in the parallel 1D geometry [see Eq. (\ref{cont})], the 
lowest-frequency mode of the continuum spectrum exceeds the plasma frequency
and grows with the magnetic field.

Up to now we considered oscillations for a string with  fixed
(strongly pinned) ends, since if the ends moved, a singular 
contribution would appear in $\varphi$. Decreasing the pinning strength, the frequency of
the localized mode decreases, and if pinning vanishes, the frequency must
vanish also. The string ends
correspond to pancakes in a layered superconductor. Thus the
localized mode is a JPR if pinning of pancakes is {\em infinitely
strong}. But for a finite pinning strength the localized mode is a vortex
mode with pancakes and strings moving together and with the frequency 
which depends on  pinning. Since pinning varies from a sample
to a sample, it is difficult to suggest an universal estimation of the frequency
of the vortex mode. But, as already discussed in Ref. \cite{Com},  the fact that
the experimentally measured $c$-axis critical current has the same dependence on
the magnetic field as the squared magnetoabsorption  resonance frequency,
confirms our suggestion that the magnetoabsorption  resonances in high fields
correspond not to the JPR, but to the vortex mode governed by pinning. 

In order to observe a vortex-mode resonance in a layered superconductor, the
inertia force proportional to the vortex mass (we neglect the Magnus force here)
should not be small compared to the friction force. A moving vortex line is a
chain of pancakes connected by random strings of average length $ r_w$
\cite{LG}, and  $ r_w$ may be considered as a size of an extended vortex
core. The Josephson length $\lambda_J=\gamma s$ ($\gamma $ is the anisotropy
ratio) usually exceeds $r_w$. The mass $M$ of the short string is determined
mostly by the vicinity of two pancakes, and in Eq. (\ref{mass}) $\partial
\varphi_{0} /\partial x =- y/r^2$ [or $(y-r_{w})/r^2$]. Then the mass per unit
length is   (cf. Ref. \cite{VM}) \begin{equation} \mu_s={M\over s}=
{\varepsilon \Phi_0^2 \over  8\pi^2 s^2c^2}  \ln{r_w\over  \xi_{ab}}~,
 \label{mus} \end{equation}
where  $\xi_{ab}$ is the coherence length in the $ab$ plane.  On the other
hand, the dissipation rate  for string motion is also proportional to the electric
energy: $\eta v_L^2 = \int { \sigma_c  E^2 }(d\vec r)_{2}$, where $\sigma_c$ is
the normal conductivity along the $c$ axis and  $\eta$ is the friction
coefficient per unit vortex length along the $c$ axis. Then the quality factor of the resonance (the ratio
of the inertia force to the friction force) is $Q=\mu_{l} \omega_{0}/
\eta=\varepsilon \omega_{0} / 4\pi \sigma_c$.
It is identical to the quality factor of the uniform plasma oscillation
as one can see comparing the supercurrent and the normal current in the
expression for the total current $ j= -({\varepsilon \omega_p^2 / 4\pi i\omega}) 
\vec E + \sigma_c  \vec E$. Thus normal-current
dissipation is not more dangerous for the vortex mode than for the plasma mode. 

The assumption of infinitely strong pancake pinning cannot be valid everywhere, 
and therefore the JPR mode cannot be an universal explanation of the
magnetoabsorption resonances. This explanation was based on  Eqs. (\ref{JPR})
and (\ref{phi}). In Refs. \cite{PRL,Repl} I argued that in order to 
derive the observed anticyclotronic dependence from these equations, 
one must make  a too
strong assumption that $r_w \sim a$ in a wide interval of fields. The 
condition  $r_w \sim a$ means a complete disintegration of vortex lines. Note
that now Bulaevskii {\em et al.} \cite{BKVM} also believe that this condition
does not take place and line structure retains in the vortex liquid. Then they
must  accept that  it is impossible to explain the anticyclotronic behavior in
the liquid phase in terms of JPR. As discussed in Ref. \cite{Com}, the line
structure in the vortex liquid was confirmed by experiments on the microwave
absorption \cite{EBT}.

The analysis given above shows that Eq. (\ref{JPR}) itself  is not always valid.
In addition, in high magnetic fields ($a \ll \lambda_J$) one may not use Eq. 
(\ref{phi}) for the decoupled state of the vortex matter in which pancakes
positions in neighboring layers are uncorrelated, but the 2D itralayer
crystalline order still exists. Then the Josephson strings form an ordered 2D
lattice in any interlayer spacing. In high magnetic fields when $\lambda_J$
exceeds $a$, Eq.  (\ref{phi}) is valid if orientations of the Josephson strings
are random, and there is no interference between contributions of distant strings
(at distances larger than $a$) to $\langle\varphi_0(\vec r)^2 \rangle$
\cite{LG}. But if the strings form a lattice, the distribution of
$\varphi_0(\vec r)$ is similar to the distribution of the current component
$j_y$ for the vortex lattice in the mixed state. This means that the
contributions from distant strings to $\varphi_0(\vec r)^{2}$ cancel, and the
logarithm   $\ln(\lambda_J/ r_w)$ in Eq.  (\ref{phi}) must be replaced by
$\ln(a/ r_w)$. Then even the assumption  $r_w \sim a$ does not help to obtain
the anticyclotronic behavior. 

In summary, we explain the
experimental dependence of magnetoabsorption resonances on a low perpendicular
magnetic field \cite{Tam,MatUnp} as the following. In very low fields, before the
frequency jump, the magnetoabsorption resonance weakly depends on a field as
was predicted for the JPR \cite{PRL}, and therefore is the JPR mode. After the
jump which apparently coincides with the vortex-matter phase transition (the
second-peak line at low temperatures), the magnetoabsorption resonance becomes
the vortex mode governed by pancake pinning. This explains the anticycltronic
dependence after the jump. The nature of the frequency jump  depends on the
structure of the vortex matter in the high-field phase which is not well
established up to now. It maybe related with a jump of $r_w$  as supposed by the
decoupling scenario of the phase transition \cite{H}. The jump of $r_w$ should
result in a jump of vortex mass, or in weakening of pinning, or both.

Any scenario of the phase transition supposes that in the high-field phase
directions of vortex lines strongly oscillate \cite{Kes},
Then even the electric field strictly parallel to the $c$ axis and to the
average direction of vortices can make some tilted segments of vortex lines to
oscillate. Thus the Lorentz-force-free geometry of the magnetoabsorption
experiments in perpendicular fields does not rule out a possibility to excite
the vortex mode.

The work was supported by the grant of the Israel Academy of Sciences and
Humanities.


\begin{references}
\bibitem{M0} M.B. Gaifullin {\em et al.}, Phys. Rev. Lett. {\bf 83}, 3928 (1999).
\bibitem{Tam} T. Shibauchi {\em et al.}, Phys. Rev. Lett. {\bf 83}, 1010 (1999).
\bibitem{MatUnp} Y. Matsuda  {\em et al.}, (unpublished).
\bibitem{KS} N.B. Kopnin {\em et al.}, Phys. Rev. Lett.  {\bf 74}, 4527 (1995).
\bibitem{BKVM} L.N. Bulaevskii {\em et al.}, cond-mat/9907462.
\bibitem{PRL} E.B. Sonin, Phys. Rev. Lett. {\bf 79}, 3732 (1997).
\bibitem{GMB}  M.Gaifullin  {\em et al.}, Phys. Rev. Lett. {\bf 81},
3551 (1998).  
\bibitem{Repl} E.B. Sonin, Phys. Rev. Lett. {\bf 81}, 3552 (1998).
\bibitem{Com} E.B. Sonin, Phys. Rev. B  {\bf 60}, 15 430 (1999).
\bibitem{Fet} A. L. Fetter and M.J. Stephen,  Phys. Rev. {\bf 168}, 475 (1968).
\bibitem{LG} A.E. Koshelev {\em et al.},   Phys. Rev. B {\bf
53}, 2786 (1996). 
\bibitem{mass}  The Josephson-string mass in low magnetic fields is
field-independent, in contrast to the Josephson-vortex mass in high fields where 
the electric fields from neighboring vortices overlap \cite{Com}.   
\bibitem{VM} E.B. Sonin  {\em et al.},  Phys. Rev. B {\bf 57}, 575 (1998).
\bibitem{EBT} H. Enriquez  {\em et al.}, Phys. 
Rev. B {\bf 58}, R14 745 (1998).
\bibitem{H} B. Horovitz, Phys. Rev. B {\bf 60}, R9939 (1999).
\bibitem{Kes} P.H. Kes  {\em et al.}, J. Phys. I France, {\bf 6}, 2327 (1996). 
\end{references}
 \end{document}